\documentclass[aps,prx,twocolumn]{revtex4-1}
\usepackage{graphicx}
\usepackage{bm}
\usepackage{amsmath}
\usepackage{esvect}
\usepackage[utf8]{inputenc} 
\usepackage[T1]{fontenc}
\usepackage{textcomp}
\usepackage{hyperref}
\usepackage{amssymb}
\usepackage{array}
\usepackage{multirow}

\DeclareUnicodeCharacter{2008}{\ensuremath{-}}
\begin{document}



\title{ Toda lattice formed in nonequilibrium steady states of SWCNT}

\author{Heeyuen Koh}
\email{heeyuen.koh@gmail.com}
\affiliation{Soft Foundry Institute, Seoul National University, 1 Gwanak-ro, Gwanak-gu, Seoul, 08826, South Korea}
\author{Shigeo Maruyama}
\affiliation{Mechanical Engineering Dept., The University of Tokyo, Hongo 3-7-1, Bunkyoku, Tokyo, Japan}

\date{\today}

\begin{abstract}
Toda lattice or FPUT chain-like dynamics have been regarded as a prerequisite for explaining the length dependence of the high thermal conductivity of low-dimensional systems at the nanoscale. In this paper, a hypothetical condition is introduced that establishes a theoretical connection between the thermal conductivity of a nanoscale low-dimensional system in nonequilibrium steady states(NESS) and the canonical motion of the equation in the Toda lattice in equilibrium. The hypothesis relies on a numerically driven coarse grained molecular dynamics(CGMD) system acquired from the trajectory data of the nonequilibrium molecular dynamics(NEMD) simulation. It models the macroscopic motion from longitudinal and flexural modulation observed in NEMD as a separate Hamiltonian in CGMD, with a perturbation term governed by an overdamping process, which is assumed to dominate during heat transfer. The Smoluchowski equation for the perturbation, which is derived from the cross correlated states between two degrees of freedom, suggests that the potential energy function induced from NESS is identical to that of the Toda Lattice. The restrictions found in the hypothetical condition are well confirmed by the data from the numerically driven coarse grained model. 

\end{abstract}

\maketitle
\section{Introduction}

Toda lattice is the one dimensional discretized chain system whose intuitively given analytical solution from the nonlinear potential energy function defines the soliton propagation from the canonical equation of motion\cite{Toda1970}. Later, the eigenvalue problem derived from the Lax pair \cite{Flaschka1974} proves the Toda lattice dynamics as an integrable system. Recently, the scope of the related physics has been extended to Dyson's Brownian motions and random matrices via approaches based on generalized hydrodynamics or generalized Gibbs ensembles\cite{Spohn2019, Spohn2020, yoshimura2020}. 

The Toda lattice dynamics specifies the set of canonical variables, which are the displacement and momentum of discretized particles connected along the chain, in the wave packet that retains the initial shape of the packet without being attenuated as it propagates. This is also what the dynamic characteristic of the FPUT chain defines, which is well known through the recurrence of the initial incidence. The extremely prolonged relaxation time, sometimes described as infinite in the FPUT chain, is a feature of both soliton dynamics. This similarity provides a theoretical ground for an exceptionally prolonged relaxation time between the modes that compose the soliton, which is regarded as a suitable characteristic of the energy carrier for divergent thermal conduction\cite{Lepri2005,henry2009anomalous,Chen2021}. 



The divergent thermal conductivity observed in nanoscale systems refers to thermal conductivity increasing along the characteristic length of the system. It has been typically observed in low dimensional systems at the nanoscale, with the proportionality of $L^{\alpha}$ or $ln(L)/L$ for one- or two-dimensional systems, respectively, where $L$ is the length of the system, and $\alpha$ is specified in the range of 0.3-0.4\cite{Chen2021,Zhang2020,Lee2017,Marconnet2013}. The nonlinear potential energy function in the Toda lattice model, whose nonlinearity may produce strong phonon-phonon scattering, is incompatible with the elongation of the phonon mean free path. Yet, the role of the wave packet in resolving the bottleneck in thermal energy propagation \cite{Shiomi2006} or in regulating phonon scattering events is a possible conjecture, including a hidden mechanism on the scale of the wave packet, possibly correlated to the length of the system.

There are a few more distinguishing traits of the divergent thermal conductivity of nanoscale systems, related to infinite relaxation time \cite{Lepri2005}.  First, hypothetically assigning an infinite relaxation time to a specific range of phonon-phonon scattering in Green-Kubo-based molecular dynamics reveals the length-dependent thermal conductivity of a polyethylene chain\cite{henry2008high,henry2009anomalous}. And recently, the trajectory data from nonequilibrium molecular dynamics (NEMD), which is converted to a numerical coarse grained model explicitly show the wave packet quantified from the cross-correlated states between two independent harmonic Hamiltonians that compose the macroscopic motion of suspended SWCNTs\cite{Koh2022}. These results strongly suggest that the evolution of thermal energy in nanoscale low dimensional system, whether it is from thermal fluctuation under cross correlated states between harmonic Hamiltonians or phonon scattering, supports soliton dynamics.  

 In the absence of formal and critical inquiry into the mechanism of soliton formation in nonequilibrium low-dimensional systems, some clues are provided by various measurements on such systems, such as the existence of the ZA mode. Elaborated studies in the state of the art experiments and simulations distinguish the dependency of the degree of freedom of the fixation\cite{Saito1992,Maruyama2003,Yamamoto2004,Zhang2005,Shiomi2006,Lindsay2009,Marconnet2013,Zhang2020} that specifically depends on the existence of ZA or flexural mode, and the rigidity of the supporting materials\cite{Barbarino2015, Fugallo2014,Nika2009}.  Recent reports on angular momentum that enhances the thermal conductivity of lattice structures or low-dimensional systems indirectly indicate that phonon scattering associated with soliton propagation can emerge when a certain degree of freedom in the flexural mode is secured\cite{Zhang2020}. There remains a need for a refined description of the dynamics of these nanoscale systems of the ZA mode induced from the nonequilibrium steady state (NESS), thereby placing them within the scope of the Toda lattice, which is defined in equilibrium. %

  For theoretical grounds that answer how thermal energy in NESS leads to the canonical variables conveying a soliton in the system, quantifying the correlated fluctuations with the flexural mode in the system's dynamics is a reasonable step for initiation, as clear traces of wave packets along the tube axis\cite{Koh2022} from such correlation are already proven in the numerically derived coarse grained system of SWCNT from NEMD data. The recent theoretical approach related to the cross correlated term with the flexural modes has reported the double partial derivatives from the Smoluchowski picture using the framework of Stochastic Thermodynamics\cite{Jarzynski2017} as the governing equation of the cross correlated quantity between harmonic Hamiltonians, validated in equilibrium\cite{Koh2021} and far from equilibrium simulations\cite{Koh2025}. Subsequently, a refinement is required to show that the cross-correlated states in NESS form soliton dynamics like the FPUT chain or the Toda lattice.
 
 This study investigates whether the potential energy function, derived from the NESS condition within the Smoluchowski framework\cite{Zwanzig1973} and stochastic thermodynamics\cite{Seifert2005,Jarzynski2017,Talkner2020} applied to cross-correlated states of two harmonic Hamiltonians, can support soliton propagation as seen in Toda lattice dynamics. The derivation of the Toda lattice upon the numerically driven CG model in NESS is introduced in II. Theoretical approach. The visualization from the nonequilibrium molecular dynamics simulation is in III. Results, and IV Discussion for the parametric studies. The paper is enclosed with V. Conclusion.  


\section{Theoretical modeling}

\subsection{System definition}

The macroscopic motion of suspended low-dimensional nanoscale systems, so-called ZA mode, is the essential condition that makes the FPUT regime dominate the heat transfer process. The macroscopic motion is composed of two harmonic motions along bending and longitudinal modulation, which are in-phase modes of flexural and longitudinal phonon modes of the structure that can be regarded as the coarse grained particle(CG) model. The modulation of the SWCNT is, therefore, identically considered with the dynamics of the CG particles with the potential energy function that is empirically given to the system as below: 

\begin{align}         
H_{\ell 0} = \frac{1}{2m} \mathbf{P}^T_{\ell}\mathbf{P}_{\ell} + \Phi_{\ell},\label{eq:eq2}\\
H_{\theta 0} = \frac{1}{2I} \mathbf{P}_{\theta}^T\mathbf{P}_{\theta} + \Phi_{\theta}, \label{eq:eq3}
\end{align}

with the set of momentum for the bond length $ \mathbf{P}_{\ell}=\left[\mathbf{P}^{1}_{\ell},..,\mathbf{P}^{N}_{\ell} \right]^T$, and the angle $\mathbf{P}_{\theta}=\left[\mathbf{P}^{1}_{\theta},..,\mathbf{P}^{N}_{\theta} \right]^T$. $T$ means transpose, and $N$ is the total number of CG particles. $m$ is mass, and $I$ is inertia of the particle. $\Phi_{\ell}$ and $\Phi_{\theta}$ are the potential energy functions. 

 In a previous study, the CG particle subject to two independent Hamiltonians for translational and rotational motion is affected by the Dzhanibekov effect, which alters the trajectory due to conservative forces \cite{Koh2025}. In previous work, the adjustment of the Dzhanibekov effect using heat diffusion process as derived from Stochastic Thermodynamics frameworks offers extremely precise replication of the nonlinear macroscopic motion of SWCNT in single beads system. 
 
 Following this enhancement, let the CG system be composed of a simple beads system that has the dynamics of the macroscopic motion of SWCNT with a temperature gradient from the different temperature environments at the two ends in Eq.(\ref{eq:eq2}) and Eq.(\ref{eq:eq3}). The simple beads system has an unknown potential energy function, and the perturbation for two reasons. One is the Dzhanibekov effect, which is caused by the coupling of rotation and translation of the beads. For the precise replication of the atomic system's motion, adjustment is unavoidable including the thermal fluctuation caused by the heat transfer along the tube. Due to the thermal expansion that differs along the temperature gradient along the tube and the thermal flutuation, the potential energy and the perturbation remain undefined at this stage.    
 
When the perturbation term for Dzhanibekov effect and thermal fluctuation in NESS is modeled as $\phi$, each Hamiltonian becomes as follows:  

\begin{align}\label{eq:CG_system}
H = H_{\ell}+H_{\theta}, \nonumber \\
H_{\ell} = H_{\ell 0 } + \phi_{\ell},\\
H_{\theta } = H_{\theta 0 } + \phi_{\theta} \nonumber.
\end{align}

$H$ is the total Hamiltonian of the CG system that represents the suspended SWCNT with a temperature gradient. The perturbation $\phi$ means the cross-correlated states between $H_{\ell}$ and $H_{\theta}$ to compensate for the abnormal coupling called Dzhanibekov effect as well as the thermal fluctuation from the evolution process in NESS. 



\subsection{Thermal fluctuation in SWCNT with Smoluchowski picture}

Given by nonequilibrium molecular dynamics simulations that have been previously conducted numerous times, the numerically deriven CG system in Eq.(\ref{eq:CG_system}) with $\phi$ is quantifiable. Applying the framework of Hamiltonian of mean force(HMF) from Stochastic Thermodynamics by Jarzynski\cite{Jarzynski2017}, the governing equation for $\phi$ can be retained as the perturbation to each CG Hamiltonian and its partition function, which is briefly derived in Appendix A. Further derivation in this subsection is identical to the result derived from the equilibrium condition in the previous work\cite{Koh2025}, but included for the integrity of this paper. 
 
The probability density function $\rho$ is from the partition function defined in Appendix A. The motion of NESS in Eq.(\ref{eq:CG_system}) are supposed to be in the evolution process as   

\begin{align}\label{eq:eq_op}
\partial_t \rho = L_{C} \rho +L_{D}\rho+ L_{S} \rho. 
\end{align}

$L_{C}$ is the Liouville operator with conservative force from potential energy, which is modified from temperature gradient, and $L_{D}$ is the drift and diffusion process. No matter how the potential energy is modified, $L_{C}$  and $L_{D}$ satisfy $\partial_t \rho =0$ with the canonical ensemble with perturbation. No factor causes each operator to be distorted by a temperature gradient.

 $L_{S}$, which is the Smoluchowski operator, is supposed to mark the overdamping process that causes the source of memory effect from cross correlation. The evolution of the state variable from the probability density function is then governed by the Smoluchowski equation as below: 
\begin{align}
\label{eqn:smolu}
\xi \frac{\partial \rho}{\partial t} = \frac{\partial }{\partial x} \left(U' \rho  \right) + \frac{1}{\beta} \frac{\partial ^2 \rho}{\partial x^2}.
\end{align}
Here, $\xi$ is the damping coefficient. $U$ is the potential function.

The overdamping process is supposed to be the rearrangement of atoms in the target system that does not generate any momentum of the group of atoms, but modifies the information of momentum directly. Substituting the partition function from the Stochastic Thermodynamic framework in Eq.(\ref{eqn:ensemble}) into Eq.(\ref{eqn:smolu}) offers the governing equation for $\phi$. Altogether with the canonical equation for coordinates and momentum variables, the equation of motion for Eq.(\ref{eq:CG_system}) becomes  

\begin{align}
\dot{\mathbf{Q}}_{j} = \frac{\partial H_{j}}{\partial \mathbf{P}_{j} }, \label{eq:eq_go1} \\
\dot{\mathbf{P}}_{j} =- \frac{\partial H_{j}}{\partial \mathbf{Q}_{j}},\label{eq:eq_go2} \\
\dot{\phi}(x_j)= \frac{\partial^2 \phi(x_j)}{\partial x^2}, \label{eq:eq_cro}
\end{align}

with $j$ which is either $\ell$ or $\theta$. $x_j$ is the set of state variable so that $x_j = (\mathbf{Q}_{j},\mathbf{P}_{j}$). The partition function at temperature $T$ is 

\begin{align}\label{eqn:ensemble}
\rho(t)  = \frac{1}{\mathcal{Z_{\lambda}}}e^{-\beta (h_s(x_j;\lambda)+\phi(x_j)) },
\end{align}

with $\beta$ = 1/kT. $k$ is boltzmann constant. $h_s$ represents the enthalpy, which is total energy that CG particle has. $\lambda$ is the external control on the total energy which is not inlucded in the scope of this work and $\phi$ is the interaction energy between the target system and the environment. Specific definition and interpretation on $\phi$ and $Z_{\lambda}$ that is noted $Z$ in later are included in Appendix A.
 
\subsection{Smoluchowski picture in NESS }
 
 When SWCNT has very strict steady states that $\partial_t \rho$ is equivalent to any location along the tube length with the linear temperature gradient from $T_c$ to $T_h$ at both ends of the tube, the Smoluchowski equation, which is overdamped Langevin equation satisfies 
 
 \begin{align}\label{eq:eq_smolu_1}
\partial_x \bigg(U'_i \rho_i +\frac{1}{\beta_i}\partial_x  \rho_i \bigg) = \partial_x \bigg(U'_{i+1} \rho_{i+1} +\frac{1}{\beta_{i+1}}\partial_x \rho_{i+1} \bigg)
 \end{align}
 with 
 \begin{align}\label{eq:eq_partition}
 \rho_i = e^{-\beta_i(H_i+\phi_i)}/Z  \text{ for } i = 1,..,N. 
 \end{align}

$i$ is the index of each particle that compose the one dimensional chain for SWCNT, and $N$ is the total number of CG particle, and the definition of $\rho$ is from Eq.(\ref{eqn:ensemble}). The steady state with a temperature gradient differs in its probability density function along the tube length, and it is noted with the subscript $i$ like $\beta_i$ for $1/kT_i$ 

 The partition function is restricted to $j=\ell$ in Eq. (\ref{eq:eq_go1}) $\sim$ Eq.(\ref{eq:eq_cro}) which is longitudinal motion, and it is the canonical ensemble. Two discrepancies follow: 1) the limited number of particles to consider the heat bath, 2) heat flux along the tube length. The thermodynamic limit of an infinite number of particles is thoroughly mentioned with finite fluctuation bath in the previous study\cite{Koh2025} that describes the bending motion in the vacuum chamber. The target system is in steady state; therefore, the motion of each part of the system is exposed to a constant temperature condition. The influence of the heat flux to the macroscopic motion of the system is included in the modified potential energy and the perturbation $\phi$. 
 

When the difference between $\rho_i$ and $\rho_{i+1}$ is presumed to be the function of the distance between two distinguished locations, and let an arbitrary polynomial be represented as below: 
\begin{align}\label{eq:eq_rho_approx} 
\rho_i = e^{\Delta} \rho_{i+1}, \\
\Delta  \approx a_0 + a_1 x + a_2 x^2 ... +a_N x^N,
\end{align}

 Eq. (\ref{eq:eq_smolu_1}) becomes 
\begin{align}\label{eq:eq_smolu_3}
\bigg( U'_i - \mathcal{A}_i \bigg) = e^{\Delta} \bigg( U'_{i+1} -  \mathcal{A}_{i+1}  \bigg), 
\end{align}

with $\mathcal{A}_i \equiv \frac{1}{\beta_i}\partial_x \rho_i$  which can be written as 
\begin{align}\label{eq:eq_smolu_2}
 \frac{1}{\beta_i}\partial_x \rho_i  = \frac{1}{\beta_i} \rho_i \cdot \bigg( -\frac{1}{Z}\frac{\partial Z}{\partial x} - \beta_i \frac{\partial}{\partial x}(H_i + \phi_i)\bigg). 
\end{align}

The influence of temperature gradient and the unknown $U_i$ in $H_i$, which is altered from thermal expansion and heat energy traversing the tube, the $\Delta$ in Eq. (\ref{eq:eq_rho_approx}) is assumped to be 
\begin{align}\label{eq:rho_Delta}
e^{\Delta} = C e^{ax}, 
\end{align}
with $a=a_1$ since the thorough nonequilibrium steady states of SWCNT, the linear component in Eq. (\ref{eq:eq_rho_approx}) is dominant and varying along $x$. The rest of the term in Eq. (\ref{eq:eq_rho_approx}) is regarded as a constant $C$. Then, Eq.(\ref{eq:eq_smolu_3}) becomes 

\begin{align}\label{eq:rho_Delta2}
Ce^{ax} = \frac{U'_i+\mathcal{A}_{i}}{U'_{i+1}+\mathcal{A}_{i+1}}.  
\end{align}

For Eq. (\ref{eq:rho_Delta2}), $U'_i+\mathcal{A}_{i}$ can be assumed to be the form proportional to $e^{ax}$. The necessary condition can be clarified with the integral of $U'_i+\mathcal{A}_{i}$  is derived as followings: 

\begin{align}\label{eq:rho_Delta3}
\int  \alpha e^{ax_i}dx = \int (U'_i+\mathcal{A}_{i})dx \\
\frac{\alpha}{a}e^{ax_i} =  kT_i\ln Z -  V_i - \phi_i +const.
\end{align}

for $H_i = U_i + V_i$ where $V_i$ is the kinetic energy of the CG system. $\alpha$ is an arbitrary given parameter. The perturbation from the overdamped process of the CG system that is numerically derived from the nonequilibrium condition becomes 

\begin{align}\label{eq:TL_phi}
\phi_i = -\frac{\alpha}{a} e^{ax_i} + kT_i\ln Z - V_i.
\end{align}

With the above equation for $\phi_i$,  Eq.(\ref{eq:rho_Delta}), which is the difference of the enthalpy at $i$ and $i+1$, should be in the form of $ax$. Thererfore, the exponential term in the above equation should be canceled out. For this condition, we can define the potential energy as below:   

\begin{align}\label{eq:eq_PE}
U = \frac{\alpha}{a} e^{ax_i} + bx +C, 
\end{align}

with an additional term $bx$ to compensate $kT_i\ln Z$ to be $ax$ in Eq.(\ref{eq:rho_Delta}) is 

\begin{align}\label{eq:eq_TL_pe2}
bx-kT\ln Z = ax. 
\end{align}

$C$ is an arbitrary parameter. Note that the system has a strictly linear temperature gradient from NESS, therefore, the $kT\ln Z$ is a linear function along the tube. 

The consequence of $H+\phi$ satisfies the first assumption on the difference bewteen $\rho_i$ and $\rho_{i+1}$ in Eq.(\ref{eq:eq_rho_approx}) as following:
\begin{align}
H_i+\phi_i = ax +const.   
\end{align} 

The overdamping process which exceeds the evolution from momentum as defined in Eq.(\ref{eq:TL_phi}), eliminates the influence of kinetic energy in total enthalphy, $H_i+\phi_i$.  Eq.(\ref{eq:eq_rho_approx}) that indicates the linear form of enthalpy along the tube length is essential to deduce the potential energy $U_i$ and the perturbed kineter energy $\phi$ to form the dual lattice that shares the same form with $e^{ax}$.

\section{Results}
\subsection{Visualization of wave packet}
The wave packet proposed from Toda Lattice formulation is from the solution of the eigenvalue problem defined with the potential energy of 

\begin{align}
U_{toda} = \frac{a_T}{b_T}(e^{b_Tx}-1)+a_Tx, 
\end{align}

with $a_T b_T>0$. $a_T$ and $b_T$ are the parameters in the Toda Lattice potential energy, which are not specified. The analytic solutions from Toda and further derivation by Fraschka\cite{Flaschka1974} are 

\begin{align}
e^{-x_i}-1 = \beta^2 \text{sech}^2 (\alpha i  \pm \beta t),\label{eq:TL_1}\\
s_i = \beta \tanh (\alpha i \pm \beta t),\label{eq:TL_2}\\
S_i = \log \cosh (\alpha i \pm \beta t) +const, \label{eq:TL_3}
\end{align}
with 
\begin{align}
\beta = \sinh \alpha. \label{eq:TL_4}
\end{align}

$i$ is the index of the particle. $s_i$ is the momentum of the particle and $S_i$ is $\int dt s_i$. 

  The thermal fluctuation as depicted in the Smoluchowski picture, responds to the macroscopic motion of the molecule and successfully visualizes the wave packet in NESS, as inferred from data acquired from NEMD under a few precedent conditions. Details of NEMD simulation and the data processing methods are included in Appendix B and the previous work\cite{Koh2025}.
  
 The potential energy derived from the numerically driven CGMD shares the term $e^{-x_i}$ in Eq.(24) with the analytic solution in Eq.(25), therefore the spatiotemporal distribution of those variables is supposed to have the wave packet patterns. The data acquired from NEMD is converted into a CG model by averaging various scales of CG particles, and these numerically driven CG models are examined in the spatiotemporal distribution to confirm the existence of wave packets. Within a specific range of CG particle scales, most of the SWCNTs listed in Table 1 exhibit multiple wave packets, or multitons, traversing the tubes. One of the result are in Fig. 1A and B with the (5,0) 80 nm case.  Another condition that could be confirmed from the spatiotemporal distribution has the opposite sign of the wave packets between the potential energy and the cross correlated energy as described in Eq.(\ref{eq:TL_phi}) and Eq.(\ref{eq:eq_PE}). The result of the potential energy in spatiotemporal distribution is included in Fig. 2. The $\pm$ sign for each variable is exactly opposite.  The parameters for overdamping process with $\xi$ or $\alpha/a$ in Eq.(\ref{eq:eq_PE}) are not specified theoretically in the paper.
 

  \begin{figure}
 \includegraphics[scale=0.5,trim={0 200 500 0},clip]{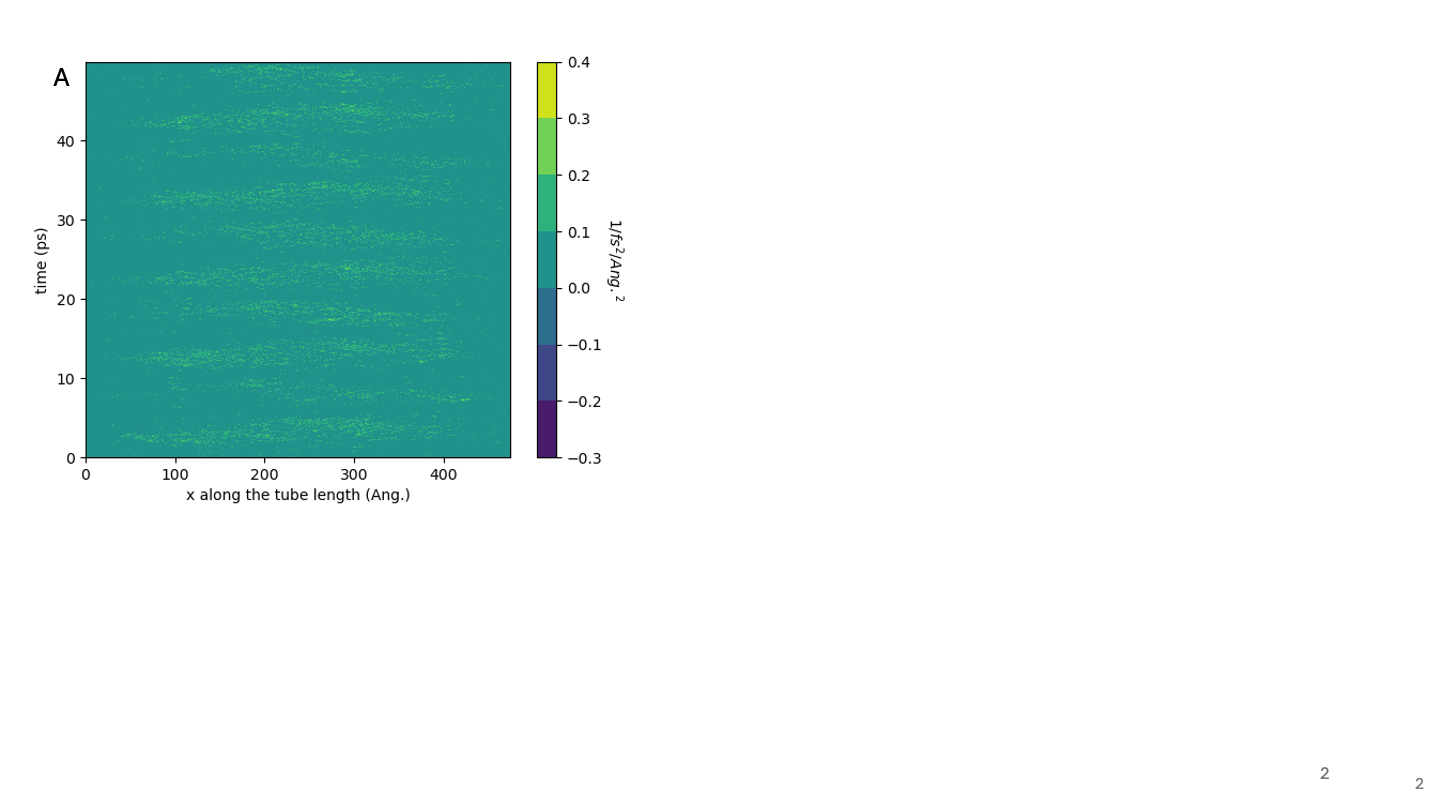}
 \includegraphics[scale=0.5,trim={0 200 500 0},clip]{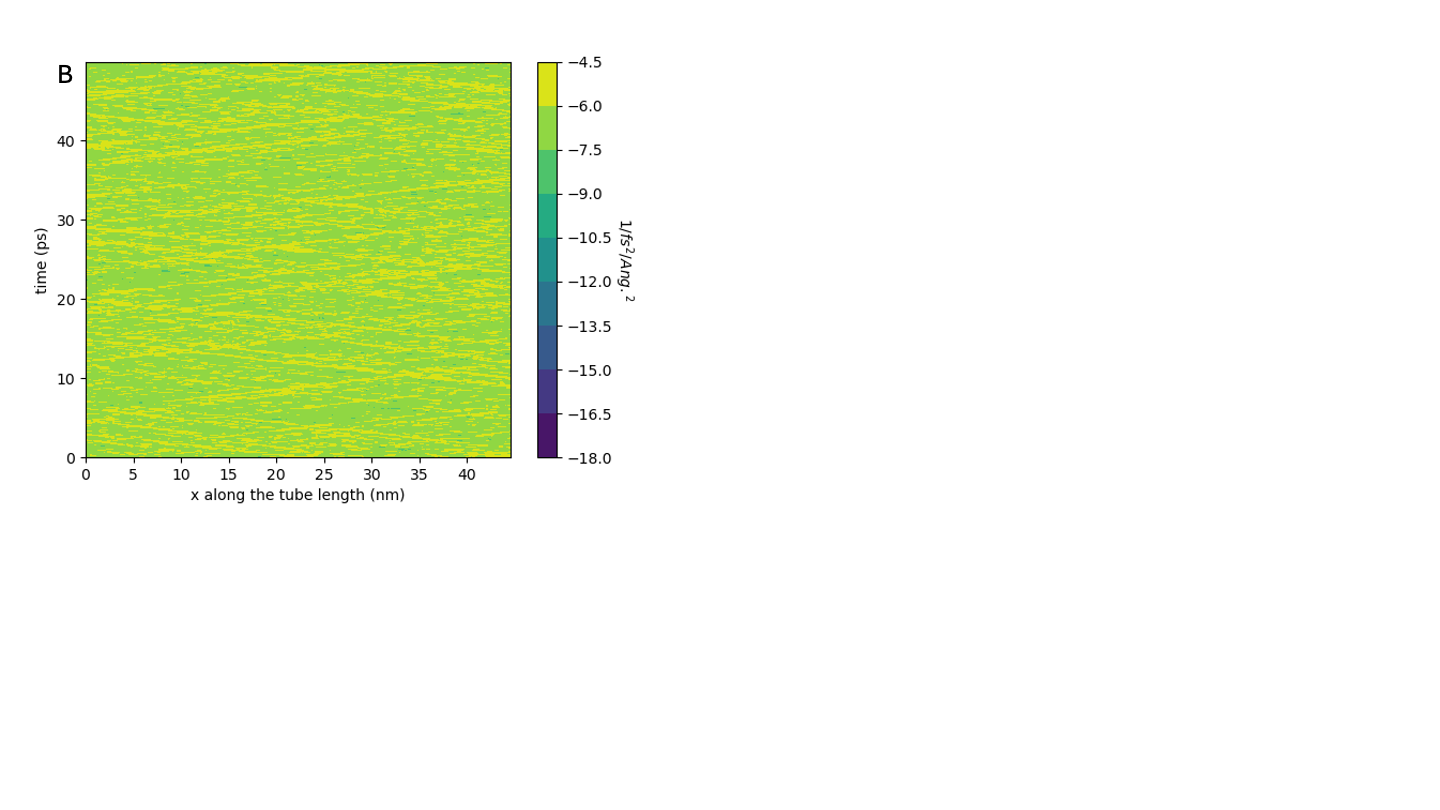}
 \caption{A. The distribution of cross correlated momentum measured in (5,5) SWCNT 80 nm model with 20 K difference at both ends with 300 K. The result of NEMD simulation is converted to the numerical CG model with a CG particle per 20 atoms. The tube axis in x and time in y. B.Spatiotemporal distribution in log scale. The same all atom NEMD simulation is averaged into simple beads model with 40 atoms as a CG particle.}
 \end{figure}

  \begin{figure}
 \includegraphics[scale=0.5,trim={0 200 500 0},clip]{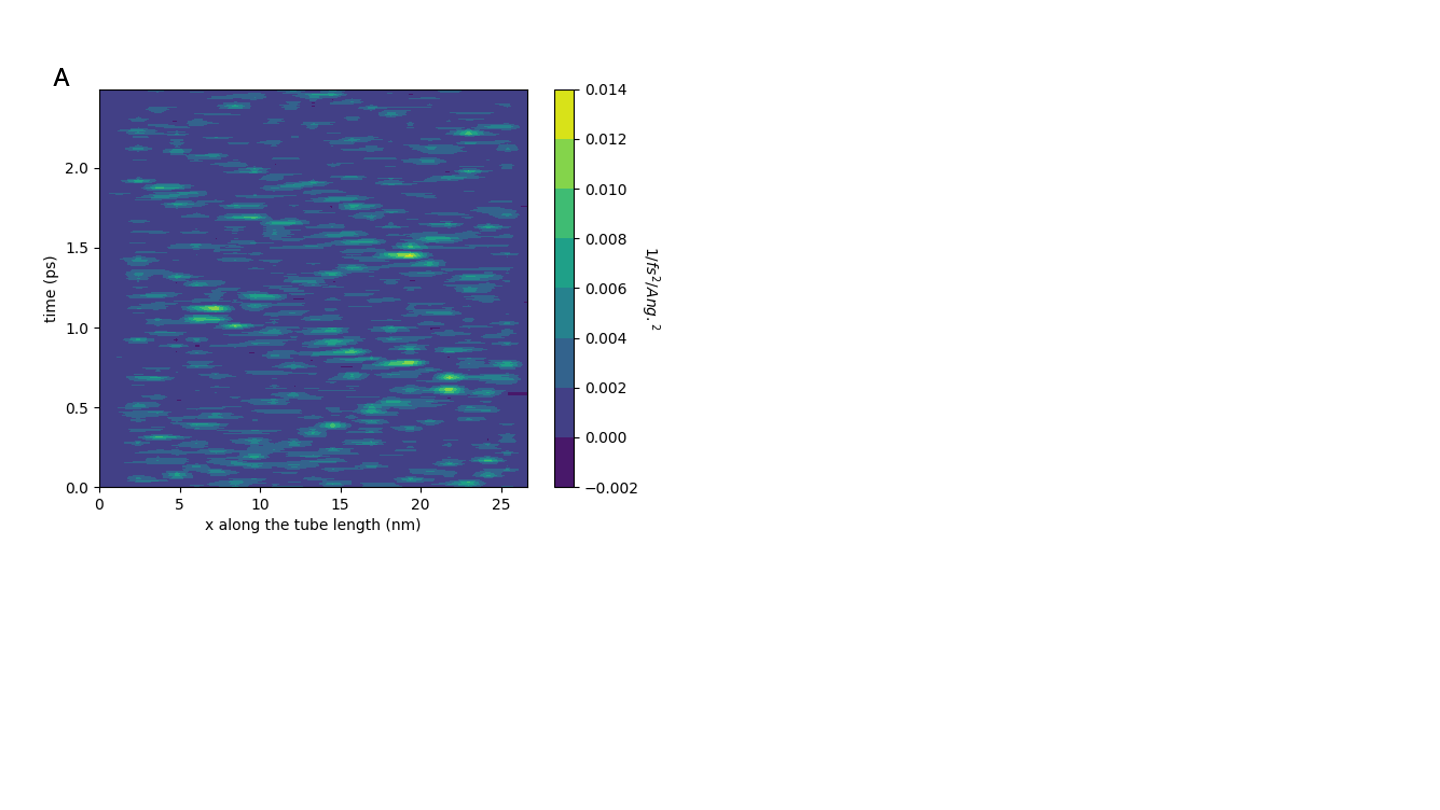}
 \includegraphics[scale=0.5,trim={0 200 500 0},clip]{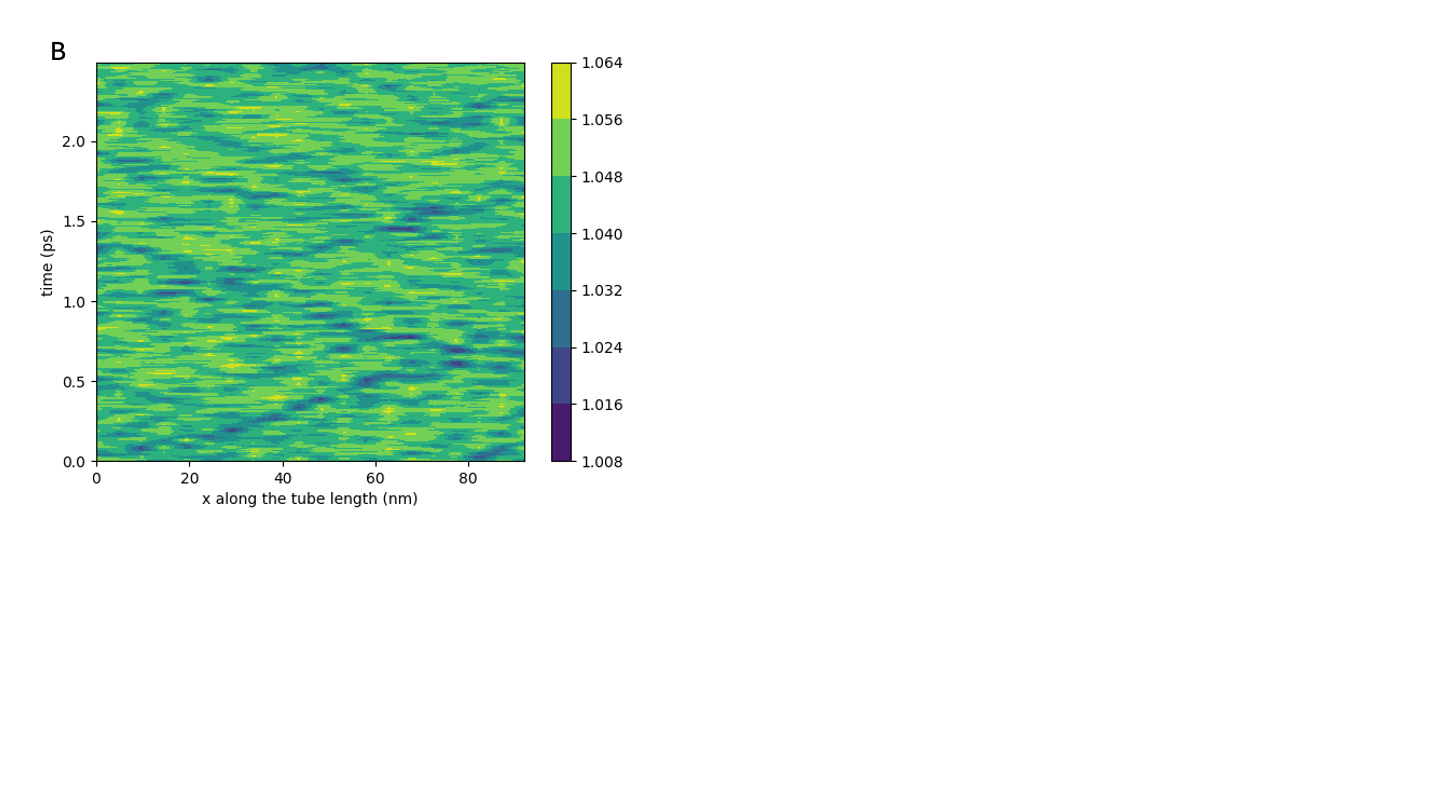}
 \caption{Wave packet distribution in (5,0) 50 nm SWCNT case with 20 K. A. The distribution of cross correlated momentum in cross correlated condition along the tube axis in x and time in y. B.Spatiotemporal distribution of potential energy $ \phi = e^{-r}+r$ calulcated using the trajectory from NEMD. Each 100 atoms is averaged for the numerical CG model.}
 \end{figure}

\subsection{Parameter study}
  
 The velocity of the wave packet $\beta / \alpha$ is proportional to the height of the wave packet. To confirm the tendency, (5,5) tube with 50 nm length is computed with two temperature difference cases. Higher temperature shows bigger amplitude with higher speed of wave packet. The result of the speed of the wave packet and its height is in Fig. X2.

\begin{table}
\centering
\begin{tabular}{ c | c | c }
\toprule
 Chiarity& Length (nm) & $\Delta$ T (K)\\
\hline
  (5,5) & 50 & 20, 40 \\
    (10,0)  & 50 &\multirow{3}{3em}{{ } {   }20}\\
     (5,0)  & 50,100 &\\
      (3,7)  &  50,100 & \\
\hline
\end{tabular}

\caption{The list of SWCNT used for analysis.}\label{t1}
\end{table}

\section{Discussion}

The environment with a temperature gradient that induces the wave packet, quantified from the trajectory data of an NEMD simulation of SWCNT using cross correlated momentum in this study, differs from that of the Toda Lattice system defined in equilibrium. However, the same potential energy can be derived using the Smoluchowski equation applied to the CG model. The existence of the wave packet, which is the soliton traversing the tube, and the following conditions during the derivation, like the opposite sign of the wave packet in the potential energy, are well proved using the data from NEMD after it is processed as a numerical CG model in the previous section. 

The reason that the overdamping process from Smoluchowski picture dominates the dynamics of the macroscopic behavior of a one dimensional system like SWCNT in NESS is from the kinetic energy in the enthalpy, $H+\phi$, which is eliminated at the final derivation of the governing equation in Eq.(\ref{eq:TL_phi}). As the perturbation $\phi$ is remained instead of the kinetic energy in the Smoluchowski equation, the overdamping process is regarded as the main governing equation for the modified potential energy and its counter part kinetic energy $\phi$ that describes the thermal fluctuation that defines the correlation between harmonic Hamiltonians.  


In the previous study, the cross correlated states are introduced as the deformation rate caused by the bending motion of SWCNT\cite{Koh2021}. However, from the point of the phonon modulation that is represented as the harmonic oscillations of the CG particle in the numerically driven CG model, the cross correlation is undoubtedly the consequence of the phonon scattering process, which is another aspect of the definition of thermal energy evolution.

 The cross correlation process derived from $\phi$ in Fig.1 and Fig. 2 is presumably from a very specific range of wavelengths or mean free paths because the observation of the wave packets and related restrictions is limited to a particular size of CG particles, and the range of the scale seems highly dependent on the chirality and aspect ratio of the tube. The theoretical explanation remains in the works on Thermodynamics theories that can count the irreversibility and entropically driven energetics related to the heat diffusion equation to explain such restrictions in scale of coarse graining.

\section{Conclusion}
Starting from the cross correlated states that make the precise replication of SWCNT's nonlinear bending motion in the previous study\cite{Koh2025, Koh2021}, the heat diffusion process derived from the Smoluchowski equation can be regarded as the proper expression for the thermal fluctuation that affects the macroscopic motion of the system. In this paper, the perturbation from the Smoluchowski equation is considered with temperature gradient under the assumption that the change of the probability density function along the tube is identical in NESS. The result is a modified potential energy function derived from the overdamping process, which is in the same form as the Toda Lattice. Then, the numerically derived CG model, derived by averaging NEMD results, confirms the pattern of the soliton traversing the tubes, as well as the few restrictions proposed during the derivation from the hypothetically suggested condition.

\bigskip

 \acknowledgements
 This research is supported by Basic Science Research Program through the National Research Foundation of Korea(NRF) funded by the Ministry of Education (NRF-2022R1I1A1A01063582). Its computational resources are from National Supercomputing Center with supercomputing resources including technical support (KSC-2020-CRE-0345).  There are no conflicts to declare.

\appendix 

\section{Hamiltonians with perturbation for cross correlated states}
Due to the adiabatic environment in the vacuum chamber, there is a very limited number of particles, which is far below being regarded as the thermodynamic limit of an infinitely large system for the definition of the partition function. The probability density function from a finite bath \cite{Campisi2009a} starts from the very beginning of its definition, that is dependent on the heat capacity as below:

\begin{align}
\rho(z,\lambda) = \frac{\Omega_B (E_{tot}-H(z,\lambda))}{\Omega_{tot}(E_{tot})}, \label{eq:eq6}\\
\lim_{C \rightarrow \infty} \rho_{C}(z;T,\lambda) = \frac{e^{-H/T}}{Z(T,\lambda)}. \label{eq:eq7}
\end{align}

 Here, $z = \left( \bf{q}, \bf{p}\right)$ is the set of state variables for the system composed of the ideal gas, and $\lambda$ is controllable parameter. The density of the state of the bath and the total system are $\Omega_{B}(E)$ and $\Omega_{tot}(E)$, respectively. The form of the probability density function, $\rho$ in Eq. (\ref{eq:eq6}) becomes as Eq. (\ref{eq:eq7}) when the specific heat $C \rightarrow \infty$ as $dn \rightarrow \infty$ where the degree of the freedom of the system is $d$ and the number of particles is $n$. 
 
A more specific condition for $  C\rightarrow \infty$ in harmonic oscillator represented is derived in the previous work\cite{Koh2025}. The result of the derivation proves that the well-known condition $dn \rightarrow \infty$, but it also shows that the finite amount of $dn$ from the number of CG particles around $\mathcal{O}(10^2) \sim \mathcal{O}(10^3)$ can afford the condition for $C \rightarrow \infty$ that overcomes the limit of the number of states even with the system defined in the adiabatic condition. 

Then, the ensemble of the system of interest can be derived from Stochastic Thermodynamics framework becomes
\begin{align}\label{eqn:ensemble}
\rho(t)  = \frac{1}{\mathcal{Z_{\lambda}}}e^{-\beta (h_s(x_j;\lambda)+\phi(x_j)) },
\end{align}
with 
\begin{align}
\phi(x_j) = \phi(x_j:N,P,T), \nonumber \\ 
=-\beta^{-1}\frac{ \int dy_j exp[\ -\beta ( H_{\mathcal{E}}(y_j) +h_{SE}  + PV_{\mathcal{E}}) ]\ }{\int dy exp[\ -\beta( H_{\mathcal{E}}(y_j) + PV_{\mathcal{E}} )]\ }, 
\end{align}
and  
\begin{align}
\mathcal{Z}_{\lambda} (N,P,T)=\int dx_j e^{-\beta (h_s+\phi(x_j))}.
\end{align}

  $x_j = \left( \bf{Q}_j, \bf{P}_j \right)$ are the state variables separately described for each system along $j=\theta$ or $\ell$. $\theta$ represents the angle and $\ell$ represents the bond length. $\mathbf{Q}_{\theta}$ and $\mathbf{Q}_{\ell}$, which are the set of vectors for the displacement of the CG particle along the bond lenth and angle wise deformation. When the state variable $x_j$ with $j=\ell$ belongs to $H_{\ell}$ then $x_{\theta}$ becomes part of $y_{\ell}=\left(\bf{Q}-\bf{Q}_{\ell},\bf{P}-\bf{P}_{\ell} \right)$. The thermal environment $\mathcal{E}$ for the Hamiltonian $H_{\theta}$ is therefore $H_{\ell}$ and connected tubes on both ends of the system. $H_{\mathcal{E}}$ is the Hamiltonian for the environment; therefore, it becomes equivalent to $H_{\theta}$ when $j=\ell$. $h_{SE}$ is the interaction energy between the target system and the environment. $PV_{\mathcal{E}}$ is the amount of energy given by the control factor. 
  
  The simplification of $\phi(x_j) $ in NPT ensemble suggests the term to be $PV_s$, the deformation energy of the conformation change of the target system $V_s$ from the control factor like the external pressure $P$, following the Stochastic Thermodynamics frameworks suggested by Jarzynski\cite{Jarzynski2017}. The detail of the derivation is included in the previous works. 
  
$\rho$, the probability density function, can be derived from Jarzynski's frameworks as the ensemble for the target system: 
\begin{align}
\rho = p^{eq} = \frac{e^{-\beta [u_s + \phi}}{\int dx e^{-\beta [u_s+\phi]}} = \frac{A}{Z}.
\end{align}

$A$ is $e^{-\beta [u_s + \phi]}$, and $Z$ is $\int dx e^{-\beta [u_s+\phi]}$. The LHS divided by $\xi$ becomes:
\begin{align}
\frac{\partial \rho}{\partial t} = \frac{A}{Z} \left( -\beta \frac{\partial}{\partial t}(u_s + \phi) \right).
\end{align}

The first term of RHS of Eq. (\ref{eqn:smolu}) is:
\begin{align}
\frac{\partial}{\partial x} \left(U' \rho \right) = \frac{\partial ^2 U}{\partial x^2} \rho + U' \frac{\partial \rho}{\partial x} \\
\frac{\partial \rho}{\partial x} = - \frac{A^2}{Z^2} -\beta \frac{A}{Z} \frac{\partial}{\partial x} (u_x + \phi).
\end{align}

The second term of RHS of Eq.(\ref{eqn:smolu}) is as below:
\begin{align}
\frac{1}{\beta} \frac{\partial^2 f}{\partial x^2} = \frac{1}{\beta} \frac{\partial}{\partial x} \left( \frac{\partial}{\partial x} \left( \frac{A}{Z}\right) \right), \\
= 2\frac{A^3}{Z^3} + \frac{A^2}{Z^2} \frac{\partial ^2 Z}{\partial x^2} + \frac{1}{Z} \frac{\partial^2 A}{\partial x^2}.
\end{align}

To make the assumption of $\frac{A}{Z}<<1$, we need the restriction on $|\frac{\partial A}{\partial x}| < 1/\beta = kT$. The restraint indicates that the possible number of states should be varied in the range of $kT$ between its neighbors. Otherwise $\frac{A}{Z}<<1$ is not satisfied so that the high order of $\frac{A}{Z}$ can not be eliminated. The full derivation of Eq.(\ref{eqn:smolu}) is as below:
\begin{align}
\xi \frac{A}{Z} \frac{\partial}{\partial t} \left(u_s + \phi \right), \nonumber \\
= \frac{\partial ^2 U}{\partial x^2} \frac{A}{Z}-\frac{1}{\beta} \left(-\beta \frac{A}{Z}\frac{\partial ^2}{\partial x^2} \left(u_s + \phi \right) \right), \nonumber \\
\frac{\partial}{\partial t}(u_s+\phi)= \frac{1}{\xi}\frac{\partial^2}{\partial x^2}\left(\phi \right). \label{eq:eq_neq}
\end{align}

$\frac{\partial u_s}{\partial t}$ is zero for equilibrium condition and even in nonequlibrium steady states. The substitution of Eq. (\ref{eqn:ensemble}) to Eq.(\ref{eqn:smolu}) is as follows:  
\begin{align}
\label{eq:smolu_eq}
 \partial_t \phi =\frac{1}{\xi} \frac{\partial ^2  \phi}{\partial x^2}.
\end{align}

\section{Simulation}
MD simulation data is adapted from Ref. \cite{Maruyama2003} where 400 nm long (5,5) SWCNT at 300 K is calculated using Brenner and aiREBO potential energy function. Time integration is conducted with 0.5 fs of time step for Velocity Verlet algorithm. Both ends are fixed with phantom wall and thermostatted with 290 K and 310K. 1 ns of NVE condition is performed before applying temperature gradient, which is conducted for 1 ns. The displacement of different number of rings in (5,5) SWCNT is averaged to compose simple beads system as shown in Fig. 2. Bond length definition is the Green-Lagrangian strain energy and angle between 3 neighboring atoms. Since the sampling frequency is 20 THz, the velocity in this analysis implies the deformation rate of the unit length, the 2 rings and 4 rings of the tube.
\bibliography{TL_SWCNT.bib}
  
\end{document}